\documentclass[twocolumn,showpacs,aps,prb,superscriptaddress,floatfix,unsortedadresss]{revtex4-1}
\usepackage{graphicx}
\usepackage{dcolumn}
\usepackage{bm}
\usepackage[utf8x]{inputenc}
\usepackage[utf8x]{inputenc}
\usepackage[autostyle]{csquotes}

\begin{document}

\author{J.J. Wagman}
\affiliation{Department of Physics and Astronomy, McMaster University, Hamilton, Ontario, L8S 4M1, Canada}

\author{D. Parshall}
\affiliation{NIST Center for Neutron Research, National Institute of Standards and Technology, Gaithersburg, Maryland 20899, USA}

\author{M. B. Stone}
\affiliation{Quantum Condensed Matter Division, Oak Ridge National Laboratory, Oak Ridge, Tennessee 37831, USA}

\author{A.T. Savici}
\affiliation{Neutron Data Analysis and Visualization Division, Oak Ridge National Laboratory, Oak Ridge, Tennessee 37831, USA}

\author{Y. Zhao}
\affiliation{NIST Center for Neutron Research, National Institute of Standards and Technology, Gaithersburg, Maryland 20899, USA}
\affiliation{Department of Materials Sciences and Engineering, University of Maryland, College Park, Maryland 20742, USA}

\author{H.A. Dabkowska}
\affiliation{Brockhouse Institute for Materials Research, McMaster University, Hamilton, Ontario, L8S 4M1, Canada}

\author{B.D. Gaulin}
\affiliation{Department of Physics and Astronomy, McMaster University, Hamilton, Ontario, L8S 4M1, Canada}
\affiliation{Brockhouse Institute for Materials Research, McMaster University, Hamilton, Ontario, L8S 4M1, Canada}
\affiliation{Canadian Institute for Advanced Research, 180 Dundas St. W., Toronto, Ontario, M5G 1Z8, Canada}

\begin{abstract}

We present time-of-flight inelastic neutron scattering measurements of $La_{1.965}Ba_{0.035}CuO_{4}$ (LBCO), a lightly doped member of the high temperature superconducting La-based cuprate family. By using time-of-flight neutron instrumentation coupled with single crystal sample rotation we obtain a four-dimensional data set (three {\bf Q} and one energy) that is both comprehensive and spans a large region of reciprocal space. Our measurements identify rich structure in the energy dependence of the highly dispersive spin excitations, which are centered at equivalent ($\frac{1}{2},\frac{1}{2}, L$) wave-vectors. These structures correlate strongly with several crossings of the spin excitations with the lightly dispersive phonons found in this system. These effects are significant and account for on the order of 25$\%$ of the total inelastic scattering for energies between $\sim$ 5 and 40meV at low $|{\bf Q}|$. Interestingly, this scattering also presents little or no $L$-dependence. As the phonons and dispersive spin excitations centred at equivalent ($\frac{1}{2},\frac{1}{2}, L$) wave-vectors are common to all members of La-based 214 copper oxides, we conclude such strong quasi-two dimensional scattering enhancements are likely to occur in all such 214 families of materials, including those concentrations corresponding to superconducting ground states. Such a phenomenon appears to be a fundamental characteristic of these materials and is potentially related to superconducting pairing.

\end{abstract}

\title{Quasi-Two Dimensional Spin and Phonon Excitations in $La_{1.965}Ba_{0.035}CuO_{4}$}

\maketitle

\section{Introduction}

The mechanism underlying high temperature superconductivity (HTS) has been intensely debated since the discovery of the first HTS, $La_{2-x}Ba_{x}CuO_{4}$ (LBCO)\cite{Bednorz_ZPhysB_1986}. Much of this research has focused on the correlation between magnetic structures and fluctuations in these systems with their superconducting ground states\cite{Timusk_RepProgPhys_1999,Fujita_JPhysSocJpn_2012,Birgeneau_JPhysSocJpn_2006,Kastner_RevModPhys_1998}. In the cuprates, this correlation is manifest as the evolution of an insulating, three dimensional, commensurate antiferromagnet to a superconducting, two dimensional (2D), incommensurate antiferromagnet with doping\cite{Keimer_PRB_1992,Wagman_PRB_2013,DoironLeyraud_Nature_2007,Stock_PRB_2008,Enoki_PRL_2013,Armitage_RevModPhys_2010,Taillefer_AnnRevCondMattPhys_2010}. This phenomenon is rich, supporting viewpoints spanning those which focus on the competition between the two ground states, to those focused on their proximity and contiguous nature \cite{Sachdev_JPhysCondMatt_2012,Kivelson_RevModPhys_2003,Lake_Nature_2002}. 

At doping levels for which commensurate antiferromagnetism is lost, the predominant magnetic excitations in La-based cuprates have a characteristic extended hour-glass dispersion, which is centered at equivalent ($\frac{1}{2},\frac{1}{2}$) positions within the pseudo-tetragonal basal plane\cite{Hayden_PRB_1996}. The dispersion approaches the commensurate position at an energy scale that is known to be dependent on the concentration of holes introduced into the copper oxide planes\cite{Matsuda_PRL_2008}, before dispersing out towards the Brillouin zone boundary at $\approx$200-300meV\cite{Coldea_PRL_2001,Headings_PRL_2010,Tranquada_Nature_2004,Hayden_PRL_1996}. Notably, these excitations display little $<L>$ dependence, indicative of two dimensional (2D) dynamic spin correlations within a three dimesional (3D) crystal structure\cite{Wagman_PRB_2013,Machida_JPhysSocJpn_1999}. 

In the last few years, a number of studies have turned to investigate the structure in the energy dependence along these dispersive modes. Such studies have appeared for $La_{2-x}Sr_{x}CuO_{4}$ (LSCO)\cite{Matsuda_PRB_2013,Christensen_PRL_2004,Lipscombe_PRL_2009,Vignolle_NaturePhys_2007}, LBCO\cite{Xu_Arxiv_2013}, as well as both Ni and Zn doped LSCO\cite{Wilson_PRB_2012,Matsuura_PRB_2012}, $YBa_{2}Cu_{3}O_{6+\delta}$\cite{Stock_PRB_2004}, and $HgBa_{2}CuO_{4}$\cite{Li_NaturePhysics_2012}. In particular, some neutron scattering studies\cite{Christensen_PRL_2004,Lipscombe_PRL_2009,Vignolle_NaturePhys_2007} on superconducting LSCO have reported striking correlations between superconductivity and a peak in the dynamic susceptibility near 20meV\cite{Lipscombe_PRL_2009,Vignolle_NaturePhys_2007}. However, all of these studies focus exclusively on scattering at the smallest {\bf Q} = ($\frac{1}{2},\frac{1}{2}, L$) 2D magnetic zone centres (2DMZCs), and do not inform on the comprehensive scattering at larger {\bf Q}, which is determined by both spin and lattice degrees of freedom, as well as any interaction between these.

Here, we report inelastic neutron scattering measurements on LBCO with x = 0.035. This sample is neither superconducting nor a 3D commensurate antiferromagnet. It displays frozen 2D incommensurate magnetic order at low temperatures, similar to LSCO of the same doping\cite{Keimer_PRB_1992,Wagman_PRB_2013}. As we will report, we find that this sample exhibits large enhancements of its highly dispersive magnetic inelastic scattered intensity at energies that correspond to crossings of the spin excitations with lightly dispersive phonons. In particular, the strongest such enhancements occur at the lowest of such spin-phonon crossings, which are near 15 meV and 19 meV in LBCO. 

\section{Experimental Details}

Single crystals were grown by the floating zone method and aligned with the $HHL$ plane horizontal\cite{Fujita_PRB_2002,Dunsiger_PRB_2008}. The sample was mounted in a closed cycle refrigerator, whose temperarture was controlled between 300 K and a base temperature of 7 K. Over this temperature range, the crystal structure for LBCO is orthorhombic with space group \textit{Bmab}\cite{Zhao_PRB_2007,Axe_PRL_1989}. However, since the mismatch between the a and b lattice parameters is slight, we will approximate the crystal structure by the high temperature tetragonal structure of LBCO, whose space group is $I4/mmm$ with a = b = 3.78$\AA$ and c = 13.2$\AA$\cite{Katano_PhysicaC_1993,Lee_JPhysSocJpn_2000}. Neutron scattering measurements were performed on the ARCS spectrometer at Oak Ridge National Laboratory using an incident energy of 60 meV. ARCS is a time-of-flight chopper spectrometer with large position-sensitive detector coverage \cite{Abernathy_RevSciInst_2012}. Coupled with single crystal sample rotation, the resulting four-dimensional (4D) neutron data set (three {\bf Q} dimensions and energy) is comprehensive and reveals the full complexity of the inelastic spectrum below $\sim$40meV for E$_{i}$ = 60 meV. Computation reduction and visualization were achieved using Mantid and Horace respectively\cite{Mantid,Horace}.

\section{Results and Discussion}
\subsection{Energy Dependent Structure to the Dispersive Magnetic Excitations}

Figure 1 shows a representative energy vs. wave-vector intensity contour at 7 K. This data, and all the data presented in this paper, have had an empty cryostat data set subtracted from them in order to better isolate the crystal signal from the background due to the cryostat. We discuss this subtraction further in the Supplemental Material\cite{supplemental}. This projection of the master 4D data set employs two integrations. The first is a narrow integration along $<H\bar{H}>$. We couple this with a fairly large integration along $<L>$. This ensures we capture the full 2D magnetic scattering, which is evident as isotropic rods along $<L>$ (see the ($\frac{1}{2},\frac{1}{2}$) rods in the panels of Fig. 3). As a result, Fig. 1 displays a series of equivalent 2DMZCs ranging from ($\frac{1}{2}$,$-\frac{1}{2}$) to ($-\frac{5}{2}$,$-\frac{7}{2}$). Note that this measurement does not resolve the incommensuration of the magnetic scattering but instead shows highly dispersive rods of inelastic scattering centered on equivalent ($\frac{1}{2}, \frac{1}{2}$) positions.

We observe clear structure in Fig. 1 as a function of energy near the 2DMZCs. Focusing on ($\frac{1}{2}, -\frac{1}{2}$), where the magnetic scattering is strongest and the phonon scattering is weakest, the scattered intensity is much greater below $\sim$19meV than above. We also see considerable enhancement of the $\sim$19meV scattering, which is pronounced at higher-$|{\bf Q}|$ equivalent 2DMZCs, such as ($-\frac{3}{2}, -\frac{5}{2}$). We find that this structure develops at the many positions in {\bf Q} and energy that correspond to the crossings of the spin excitations with phonons. - particularily, the crossing where the low-lying $\sim$19 meV optic phonon crosses the 2DMZCs\cite{Axe_PRL_1989}.

\begin{figure}
\centering 
\includegraphics[scale=0.2]{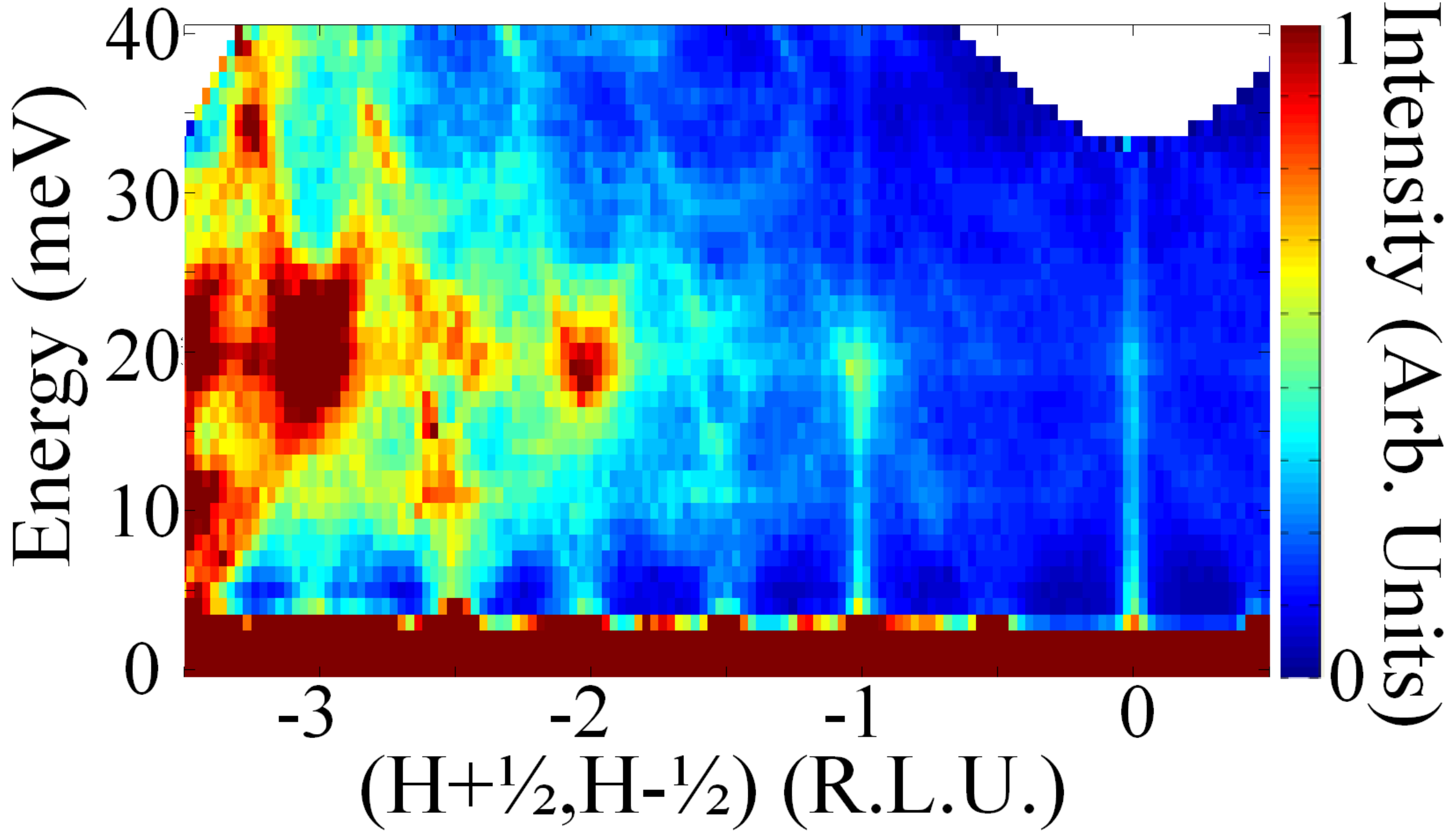} 
\caption{Energy vs. wave-vector neutron intensity map for $La_{1.965}Ba_{0.035}CuO_{4}$ showing the scattered neutron intensity along (H+$\frac{1}{2}$, H-$\frac{1}{2}$) at 7 K. The data employs a subtraction of an empty cryostat data set, integration from 0.4 to 0.6 R.L.U. in $<H\bar{H}>$ and -4 to 4 R.L.U. in $<L>$, where R.L.U. means in units of the reciprocal lattice. The vertical rod-like features, emanating from ($\frac{1}{2}$,$-\frac{1}{2}$) and equivalent 2D magnetic zone centres, are the dispersive magnetic excitations.}
\end{figure}

At higher $|{\bf  Q}|$, we find that this crossing enhancement grows dramatically in intensity and breadth in both energy and width along {\bf Q}. While purely magnetic scattering drops off with increasing $|{\bf Q}|$ as the square of the magnetic form factor, phonon intensities scale roughly as the magnitude of {\bf Q} squared\cite{Squires_Text}. Indeed, we see in Fig 1. that away from the spin-phonon crossings, the dispersive rods of magnetic excitations diminish with increasing $|{\bf Q}|$, while the phonon-dominated non-magnetic background increases markedly with $|{\bf Q}|$. Therefore, the large enhancement of the (-$\frac{3}{2}, -\frac{5}{2}$) and (-$\frac{5}{2}, -\frac{7}{2}$) scattering near $\sim$19 meV cannot be purely magnetic in origin. The enhancement is also surprisingly strong, given the modest phonon intensities at wave-vectors somewhat removed from these 2DMZCs, suggestive that the optic phonons near 2DMZCs qualitatively differ from those away from these wave-vectors.

\subsection{Scattering within the $HK$ and $HHL$ Planes}

\begin{figure}
\centering 
\includegraphics[scale=0.14]{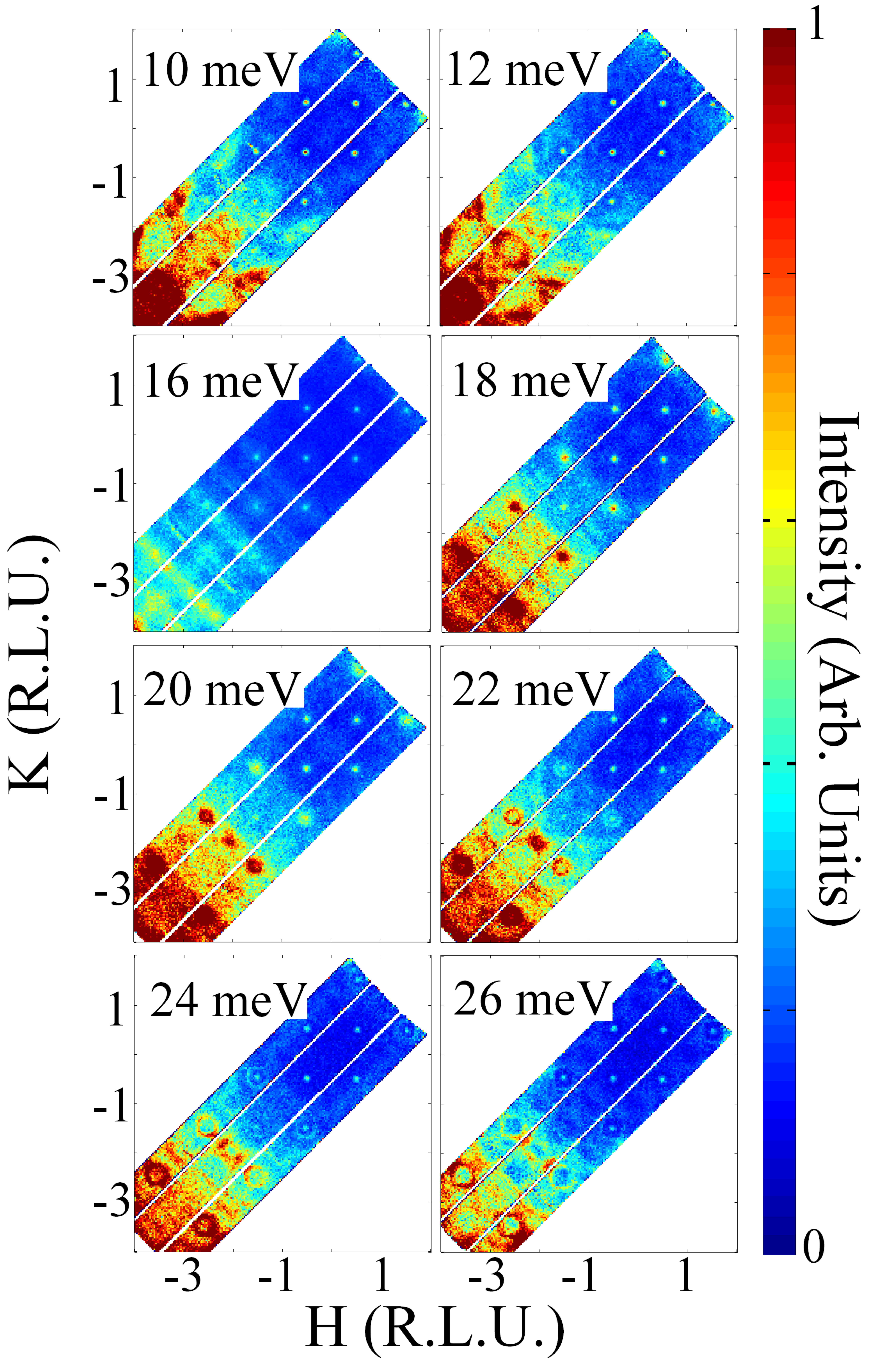} 
\caption{$HK$ reciprocal space maps of the inelastic neutron scattering from $La_{1.965}Ba_{0.035}CuO_{4}$, showing the scattered neutron intensity in the $HK$ plane at 7 K. The data employs a subtraction of an empty cryostat data set, integration from -4 to 4 in $<L>$ and an integration of $\pm$ 1 meV in energy about each listed energy. From this projection, the strong spot-shaped features at 10 and 12 meV, centered at equivalent ($\frac{1}{2},\frac{1}{2}$) positions, identify the purely magnetic scattering.}
\end{figure}

\begin{figure}
\centering 
\includegraphics[scale=0.125]{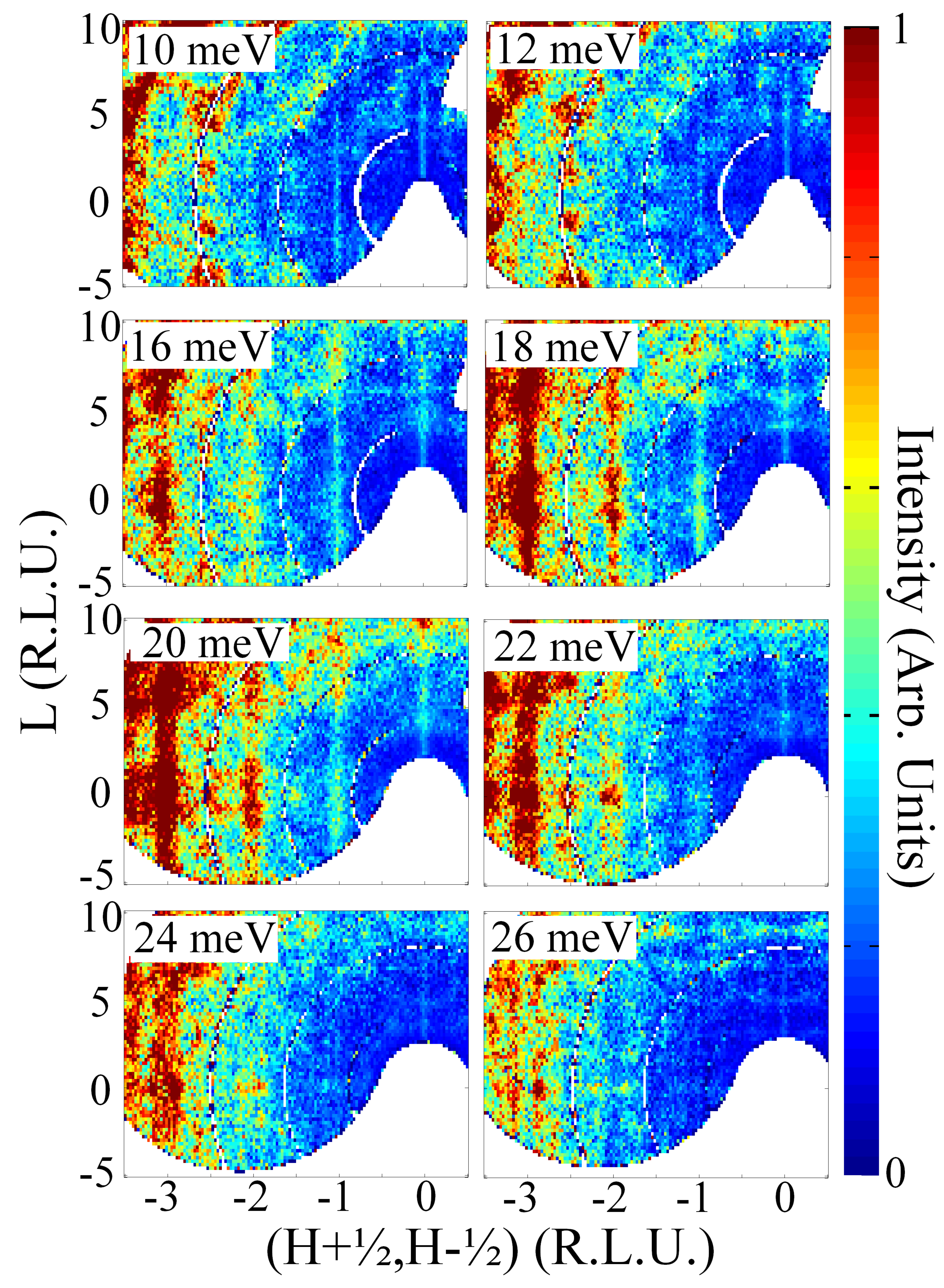} 
\caption{$(H+0.5,H-0.5,L)$ reciprocal space maps of the inelastic neutron scattering from $La_{1.965}Ba_{0.035}CuO_{4}$ showing the scattered neutron intensity in the $HHL$ plane at 7 K. The data employs a subtraction of an empty cryostat data set, integration from -0.1 to 0.1 in $<H\bar{H}>$ and an integration of $\pm$ 1 meV in energy. From this projection, the vertical, resolution-limited rods of scattering at 10 and 12 meV about ($\frac{1}{2}$,$-\frac{1}{2}$) and ($-\frac{1}{2}$,$-\frac{3}{2}$) identify the purely magnetic scattering. As the phonons begin to contribute at the higher spin-phonon crossing energies, the scattering increases substantially in both breadth and intensity.}
\end{figure}

We take advantage of the 4D nature of our data to project into another scattering plane. In Fig. 2, we integrate over the same $L$ range as in Fig. 1, but instead we now integrate by $\pm$1 meV in energy to view the scattering in the $HK$ plane. Beginning with the $HK$ maps at 10 and 12 meV (top row of Fig. 2), we can see intense spots of scattering at equivalent 2DMZCs. Again, note the drop in intensity at equivalent positions with higher $|{\bf Q}|$. Then, by 16 meV, the scattered intensity increases at these 2DMZCs. We emphasize that the scattered intensity at equivalent magnetic wave-vectors with higher $|{\bf Q}|$ is now distinguishable from background unlike at lower energies, which do not correspond to the crossing of spin excitations with phonons. At 18 and 20 meV, the scattering at all magnetic positions has increased dramatically in both intensity and breadth in the $HK$ plane - particularily at higher $|{\bf Q}|$. By 22 meV, clear ring-like excitations develop centered on the 2DMZCs. We identify these rings as arising from optic phonons, which disperse upwards in energy from a minimum near 19 meV at the equivalent ($\frac{1}{2}$,$\frac{1}{2}$) positions. The scattering from these phonons is strongest at high $|{\bf Q}|$ and is difficult to detect at low $|{\bf Q}|$. Moreover, by 22 meV, the scattered intensity at equivalent 2DMZCs has decreased and is again strongest at low $|{\bf Q}|$.

We can also explicitly look at the $L$-dependence of the scattering at equivalent 2DMZCs. Fig. 3 shows another integration of our 4D data set in energy and $<H\bar{H}>$ such that we view reciprocal space maps within the $HHL$ plane. At energies below the spin-phonon crossing, such as 10 and 12 meV, magnetic rods of scattering, indicative of 2D correlations, are identified only at low $|{\bf Q}|$ ($\frac{1}{2}, -\frac{1}{2}, L$) and ($-\frac{1}{2}, -\frac{3}{2}, L$). Between 16-20 meV, these rods of scattering become stronger and broader within the $HK$ plane and additional rod-like features appear at equivalent large $|{\bf Q}|$ positions, such as ($-\frac{3}{2}, -\frac{5}{2}, L$), similar to the trend seen in Fig. 2. These results show that the inelastic scattering at 2DMZCs between 16 and 20 meV presents as extended rods of scattering along $L$, at both relatively small $|{\bf Q}|$ positions such as ($\frac{1}{2}, -\frac{1}{2}, L$), where it would be expected to be magnetic in origin, and at relatively high $|{\bf Q}|$ positions such as ($-\frac{5}{2}, -\frac{7}{2}, L$), where it would be expected to be mainly due to phonons. This suggests a quasi-2D composite spin-phonon excitation, wherein the relevant phonon eigenvectors would couple strongly to the magnetism in LBCO.

\subsection{Quantitative Analysis of the Spin-Phonon Crossings and Their Temperature Dependence}

Figure 4, shows the temperature dependence of the energy vs. wave-vector maps over a low $|{\bf Q}|$ subset of the map shown in Fig. 1. Data is shown at three temperatures: 7, 100 and 300 K, located in panels a)-c) respectively. These three data sets have been normalized so as to be on the same intensity scale. Interestingly, the structure of the dynamics does not change significantly upon raising temperature, suggesting that they are independent of the 2D IC ordering temperatures in the system. That said, it is also clear that the scattered intensity increases markedly with temperature. However, this is, to some extent, an expected result, emanating from the temperature dependence of the Bose factor. As will become clear, a quantitative analysis is required to understand the the temperature dependence of the observed spin fluctuations.

A quantitative analysis of the inelastic scattering at and near the 2DMZCs requires a reliable background estimate. We consider two approaches to this issue. One is to estimate the inelastic background from the inelastic scattering at low temperatures and low $|{\bf Q}|$, both below the dome of scattering formed by the acoustic phonons, as well as away from the 2DMZC. This background estimate should be largely $|{\bf Q}|$ and energy independent and can be subtracted off of the scattered intensity measured as a function of energy at the 2DMZC. Another approach is to use the intensity measured within the basal plane away from the 2DMZC as an energy-dependent background. This energy dependent background method allows us to remove the background as well as any phonon contribution to the scattering with little dispersion and little $|{\bf Q}|$ dependence to their intensity. We relegate the results of this energy-dependent background analysis to our supplemental materials\cite{supplemental}. In both methods, we employ an additional set of integrations, which instead of displaying slices of the data, as those shown in Figs. 1-3, yield effective constant energy and constant-{\bf Q} cuts through the data. 

\begin{figure}
\centering 
\includegraphics[scale=0.165]{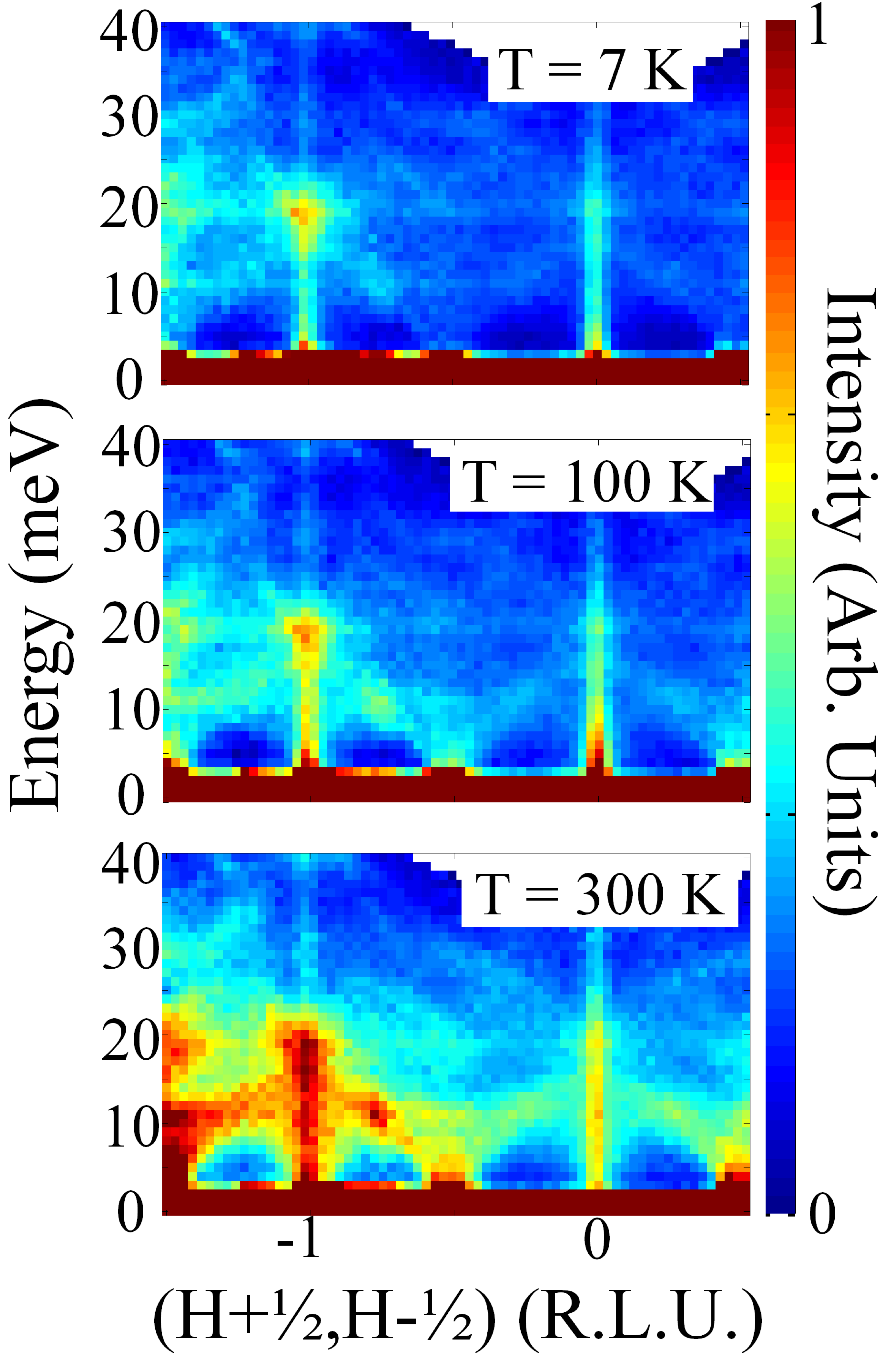} 
\caption{Energy vs. Wave-vector maps at a) 7, b) 100 and c) 300 K. $L$ has been integrated from -4 to 4 and $H\bar{H}$ has been integrated from 0.4 to 0.6, as in Fig. 1.}
\end{figure}

We now analyse the inelastic scattering at the 2DMZC using the background analysis that we described at the beginning of this subsection. As can be seen in Fig. 4, surrounding the ($\pm\frac{1}{2},\pm\frac{1}{2}$) peaks are regions of low scattering intensity, which are bounded by the nearby acoustic phonons. This scattering is largely comprised of incoherent scattering and the sample independent experimental background not captured by the empty cryostat subtraction. We can measure the total background scattering in these four regions by first integrating from -4 $\leq L \leq$ 4 and -0.1 $\leq \bar{H}H \leq$ 0.1. To then obtain the background scattering in these four background regions we further integrate in $HH$ from $\pm$0.2 to $\pm$0.35 and $\pm$0.6 to $\pm$0.8. This yields the total background scattering in these four regions, which we note are quantitatively similar to each other. Given this similarity, we average the results from these four integrations together, and make the approximation that this background is a constant for all energies and {\bf Q}. Our inelastic scattering signal at the 2DMZC is then given by effective constant-{\bf Q} scans obtained by integrating from -4 to 4 in $L$, -0.1 to 0.1 in $\bar{H}H$ and from $\pm$0.4 to $\pm$0.6 in $HH$, which we average over the four ($\pm\frac{1}{2},\pm\frac{1}{2}$) positions and display in Fig. 5 a).

\begin{figure}
\centering 
\includegraphics[scale=0.3]{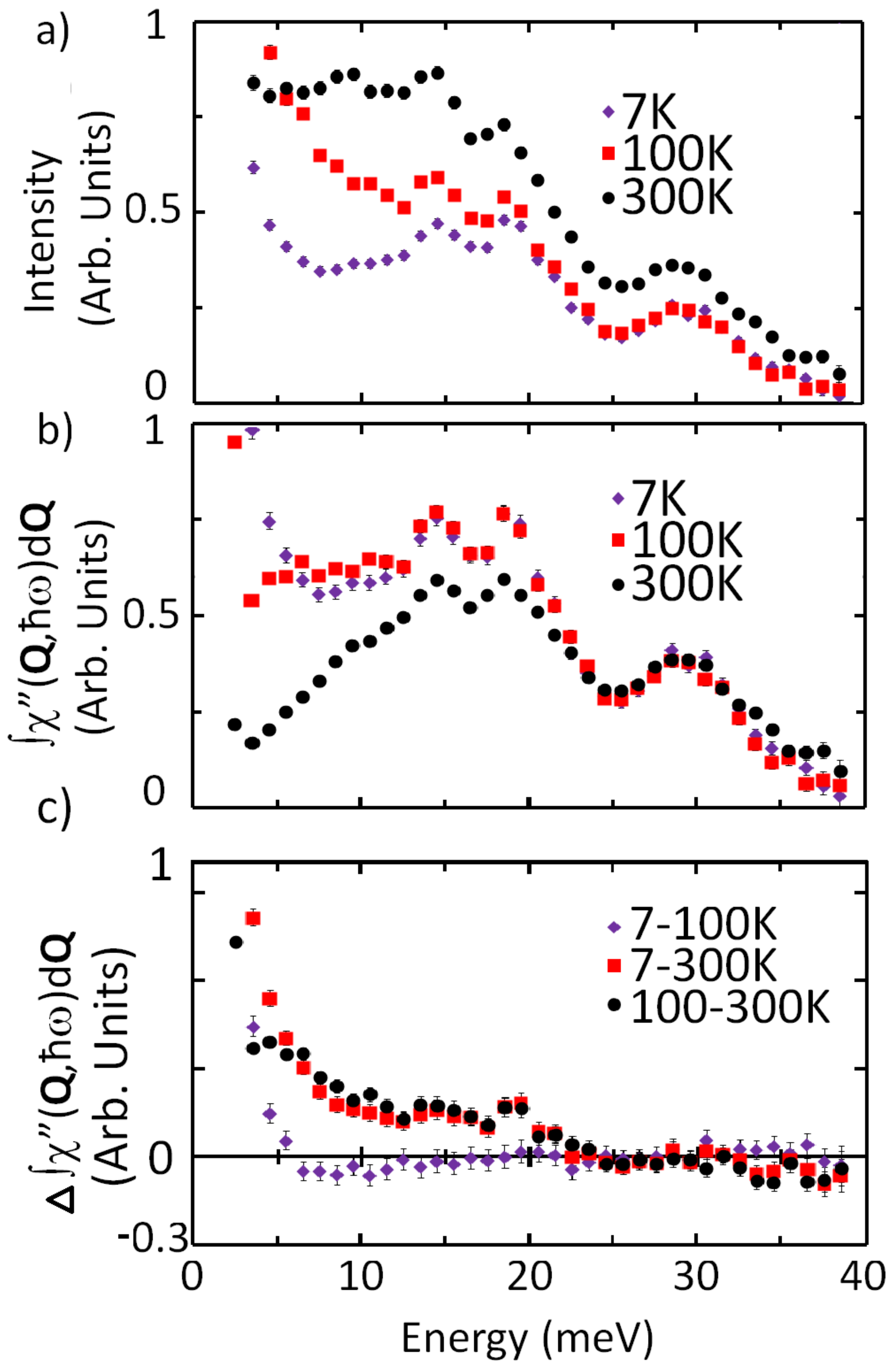} 
\caption{a) Scattered intensity averaged between areas surrounding ($\frac{1}{2},\frac{1}{2},L$) and ($-\frac{1}{2},-\frac{1}{2}$) shown as a function of energy at T = 7, 300 and 300 K. The data shown is obtained by integrating from -4 to 4 in $L$, -0.1 to 0.1 in $H\bar{H}$ and 0.4 to 0.6 in $HH$. Data from the averaged regions found by integrating $\pm$0.2 to $\pm$0.35 in $HH$ and $\pm$0.6 to $\pm$0.8 with -4 $\leq L \leq$ 4 and -0.1 $\leq \bar{H}H$ 0.1 has been used as a background. b) $\chi\prime\prime$($\hbar\omega$,{\bf Q},T) obtained from a) is also shown for T = 7, 100 and 300 K. c) Difference between 7 and 100 K, 7 and 300 K and 100 and 300 K data sets shown in b). Error bars represent one standard deviation.} 
\end{figure}

\begin{figure}
\centering 
\includegraphics[scale=0.3]{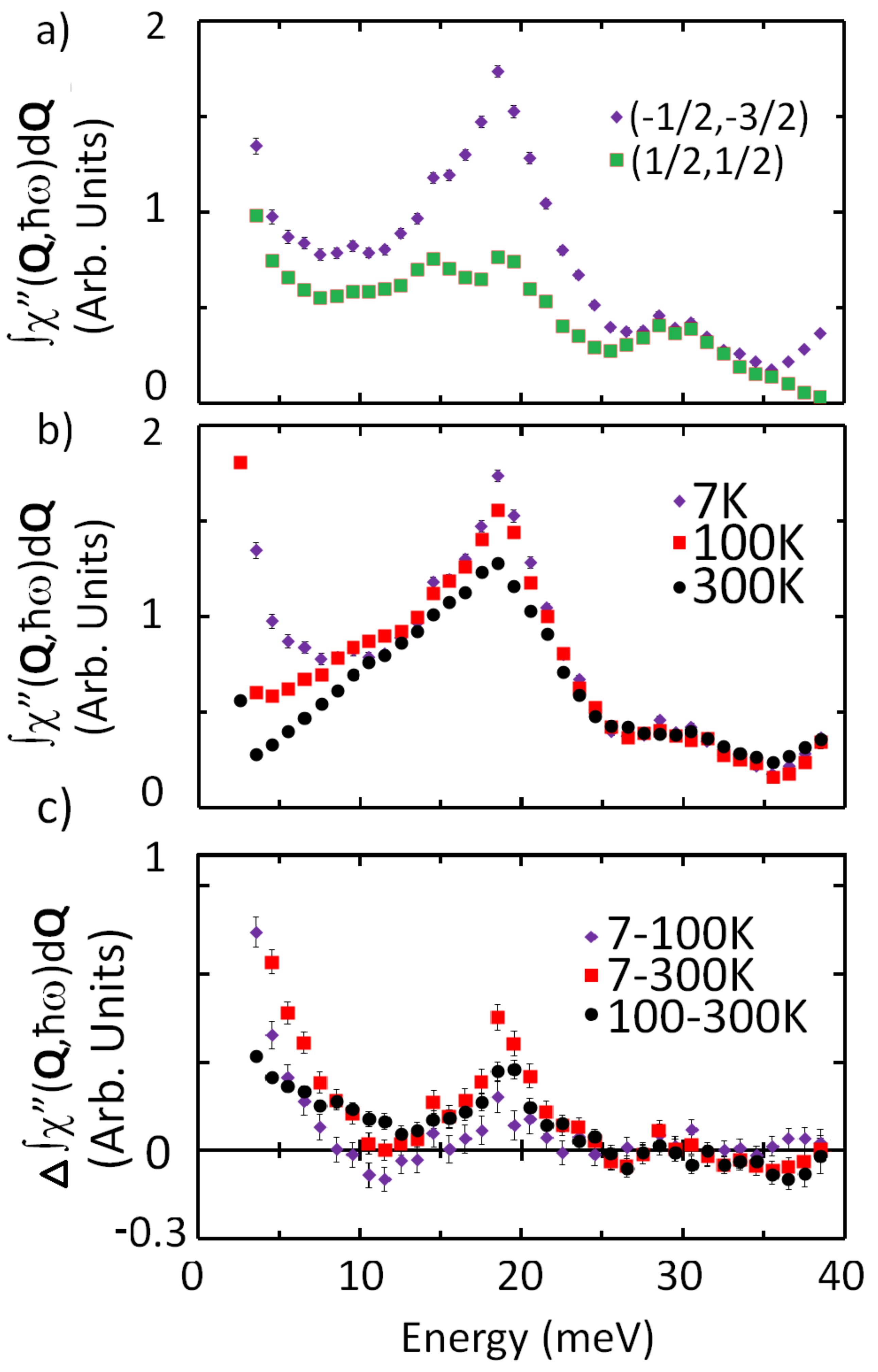} 
\caption{a) $\chi\prime\prime$($\hbar\omega$,{\bf Q},T) at both {\bf Q} = ($\frac{1}{2},\frac{1}{2},L$), which is also presented in Fig. 6 a), and ($-\frac{1}{2},-\frac{3}{2},L$) shown as a function of energy. The data shown is obtained by integrating from -4 to 4 in $L$, -0.1 to 0.1 in $H\bar{H}$ and 0.4 to 0.6 in $HH$. Data from the averaged regions found by integrating $\pm$0.2 to $\pm$0.35 in $HH$ and $\pm$0.6 to $\pm$0.8 with -4 $\leq L \leq$ 4 and -0.1 $\leq \bar{H}H$ 0.1 has been used as a background. The data has been normalized to the same scale as that used in Fig. 5 b). b) $\chi\prime\prime$($\hbar\omega$,{\bf Q},T) obtained from a) is also shown for T = 7, 100 and 300 K. c) Difference between 7 and 100 K, 7 and 300 K and 100 and 300 K data sets shown in b). This intensity scale is the same as that used in Fig. 5 c). Error bars represent one standard deviation.}
\end{figure}

The aforementioned background is then subtracted from this signal to produce the inelastic scattering function, S(${\bf Q}, \hbar\omega$) shown in Fig. 5 a) for 7, 100, and 300 K. These three temperatures correspond, respectively, to a temperature within the 2D IC AF frozen magnetic state, a temperature above any 2D IC AF magnetic transitions relevant to any doping x and representative of the ``pseudogap" phase, and a temperature relevant to 3D C AF magnetic ordering of the parent compound with x = 0, respectively. S({\bf Q},$\hbar\omega$,T) is itself given by the product of the Bose thermal population factor, $n(\hbar\omega + 1)$, which is an analytic function of the ratio of $\hbar\omega$ to temperature and enforces detailed balance, and the imaginary part of the dynamic susceptibility, $\chi\prime\prime({\bf Q},\hbar\omega,T)$. This latter function is the energy-absorbing part of the dynamic susceptibility. It is an odd function of $\hbar\omega$ and contains all the physics of the system of interest. Explicitly,

\begin{equation}
S({\bf Q},\omega,T) = [n(\hbar\omega)+1)]\times\chi\prime\prime({\bf Q},\omega,T)
\end{equation}

where

\begin{equation}
[n(\hbar\omega) + 1)] = \frac{1}{1-e^{-\frac{\hbar\omega}{k_{B}T}}}
\end{equation}

With a robust estimate for the background, and knowing the temperature, it is then straightforward to isolate $\chi\prime\prime({\bf Q},\hbar\omega,T)$. In Fig. 5 b) we show the integral in $\bf{Q}$ around $\bf{Q}$ = ($\frac{1}{2},\frac{1}{2},L$), with the same limits of integration as decribed above for S({\bf Q},$\hbar\omega$,T). Fig. 5 b) then displays this integral of $\chi\prime\prime({\bf Q},\hbar\omega,T)$ as a function of energy while Fig. 5 c) shows the difference between this integral of $\chi\prime\prime({\bf Q},\hbar\omega,T)$ for T = 7 and 100 K, for T = 7 and 300 K and for 100 and 300 K. These results show that the strong enhancement in the {\bf Q}-integral of S({\bf Q},$\hbar\omega$) at $\sim$15 meV and 19 meV is also seen in the {\bf Q}-integral of $\chi\prime\prime({\bf Q},\hbar\omega)$ around $\bf{Q}$ = ($\frac{1}{2},\frac{1}{2},L$) at T = 7 K.

A similar analysis was also carried out for the $\bf{Q}$ = (-$\frac{1}{2},-\frac{3}{2},L$) 2DMZC, using the same background as for $\bf{Q}$=($\frac{1}{2},\frac{1}{2},L$). This is shown in Fig. 6, where Fig. 6 a) compares the relevant {\bf Q}-integrated $\chi\prime\prime({\bf Q},\hbar\omega)$ from $\bf{Q}$ = ($\frac{1}{2},\frac{1}{2},L$) and $\bf{Q}$ = (-$\frac{1}{2},-\frac{3}{2},L$). We find that the same enhancement of the inelastic scattering occurs near 15 and 19 meV at T = 7 K at both equivalent 2DMZCs, although the 15 meV enhancement is difficult to resolve when compared to the 19 meV enhancement at (-$\frac{1}{2},-\frac{3}{2},L$). Similar to Figs. 5 b) and c), Fig. 6 b) shows the temperature dependence of the ${\bf Q}$-integrated $\chi\prime\prime({\bf Q},\hbar\omega,T)$ at $\bf{Q}$=(-$\frac{1}{2},-\frac{3}{2},L$) as a function of energy while Fig. 6 c) shows the difference between this integral of $\chi\prime\prime({\bf Q},\hbar\omega,T)$ for T = 7 and 100 K, for T = 7 and 300 K and for 100 and 300 K.

\begin{figure}
\centering 
\includegraphics[scale=0.33]{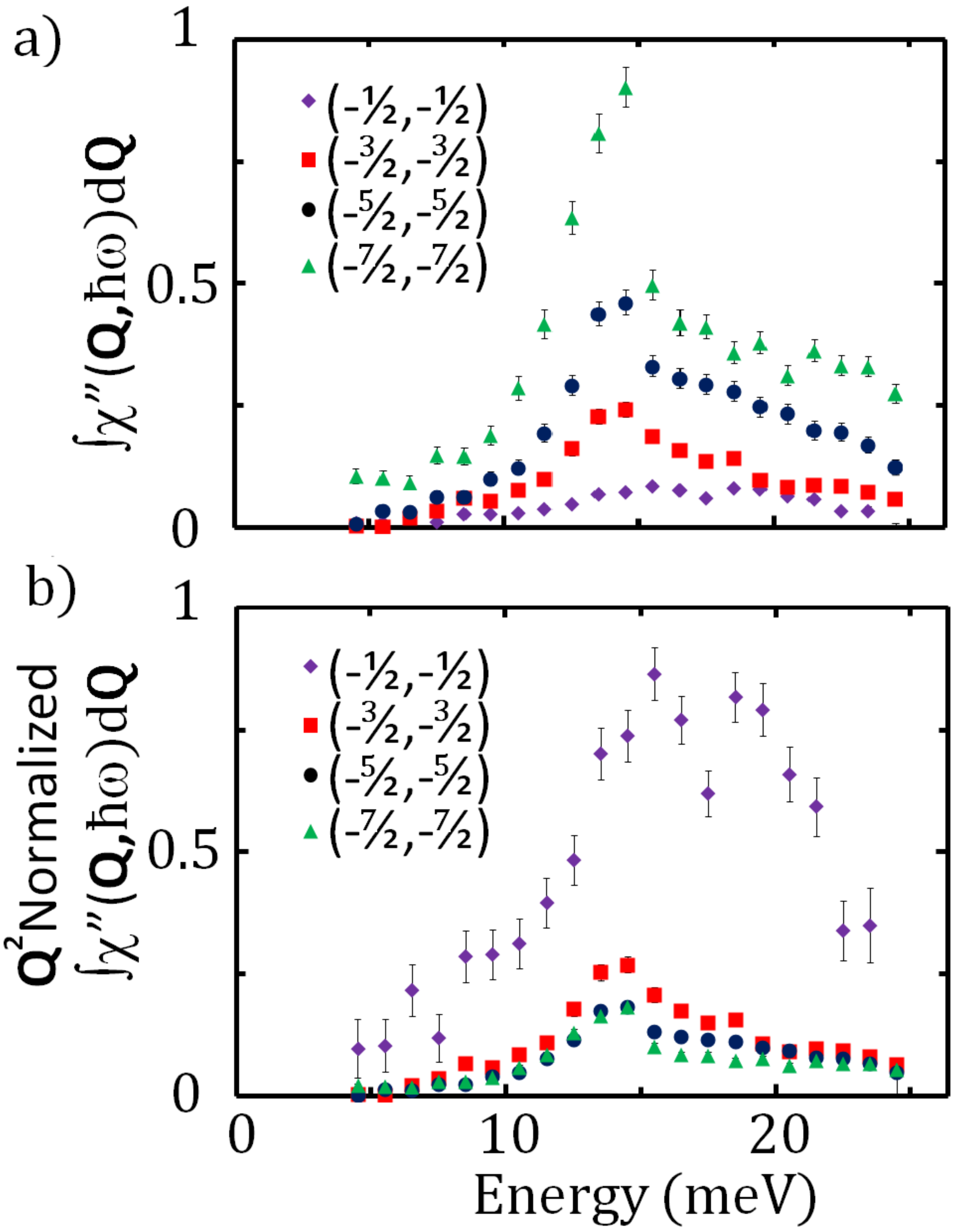} 
\caption{a) Constant energy cuts of the data using the same $HH$ and $H\bar{H}$ integrations as Figs. 5 and 6, but that also use a very small $L$ integration range. This minimizes contributions from phonon scattering, especially at the lowest $|{\bf Q}|$ positions shown. A spin-wave-like background has been subtracted from these data sets and data has also been corrected for the Bose factor, as described in the text. b) The same constant energy cuts as shown in a), but with each data set normalized by $|{\bf Q}|^{2}$. Error bars represent one standard deviation.}
\label{small_L}
\end{figure}

The temperature dependence of the integral of $\chi\prime\prime({\bf Q},\hbar\omega,T)$ at the $\bf{Q}$=($\frac{1}{2},\frac{1}{2},L$) 2DMZC shown in Fig. 5 is striking, as it shows that all of the difference between this integral at 7 K and 100 K is below $\sim$ 7 meV, while that between either 7 or 100 K and 300 K is below $\sim$ 20 meV. The loss of spectal weight at the 2DMZC on going from the frozen 2D IC magnetic state to the ``pseudo-gap" state at 100 K is at low energies, while most of the spectal weight between 7 and 20 meV remains unchanged. Instead, this 7-20 meV spectral weight diminishs on a temperature scale set by T$_N$ relevant to the undoped parent compound of LBCO with x = 0. A similar phenomenology is associated with the integral of $\chi\prime\prime({\bf Q},\hbar\omega,T)$ at the $\bf{Q}$=(-$\frac{1}{2},-\frac{3}{2},L$) 2DMZC shown in Fig. 6, although the temperature dependent spectral weight is concentrated more in the 15 - 20 meV regime.

We conclude this section by examining integrals of $\chi\prime\prime({\bf Q},\hbar\omega)$ at wavevectors of the form ($\frac{n}{2}, \frac{n}{2}, 0$) with n odd. These are the 2DMZCs such as $\bf{Q}$=($\frac{1}{2},\frac{1}{2},0$) etc. The significance of these 2DMZCs, with $L$ = 0 is that the structure factor for all wave-vectors of the form ($\frac{n}{2}, \frac{n}{2}, 0$) are identical within the $I4/mmm$ space group. Thus, the inelastic scattering at these positions should scale only as $|{\bf Q}|^{2}$ if it is due solely to one phonon creation processes. 

Fig. 7 a) shows the integral of $\chi\prime\prime({\bf Q},\hbar\omega)$ at such wave-vectors of the form ($\frac{n}{2}, \frac{n}{2}, L$), over a small range of $L$ about 0, namely -0.5 $\leq L \leq$ 0.5. This integral of $\chi\prime\prime({\bf Q},\hbar\omega)$ has been corrected for magnetic inelastic scattering at the ($\frac{n}{2}, \frac{n}{2}, 0$) positions (whose intensity does not scale as $|{\bf Q}|^{2}$), by fitting the scattering at low energies (less than 10 meV) and assuming that this magnetic strength falls off as $\hbar\omega ^{-1}$ as is expected for spin waves\cite{Tranquada_Text}. This correction may underestimate the magnetic contibution at low energies, but the net effect is to allow $\chi\prime\prime({\bf Q},\hbar\omega)$ for the 2DMZC with the smallest $|{\bf Q}|$, ($-\frac{1}{2},-\frac{1}{2}, 0$), to go to $\sim$ zero at low energies, as is expected in the absence of magnetic scattering.  

Fig. 7 b) shows the same integral of $\chi\prime\prime({\bf Q},\hbar\omega)$ as shown in Fig. 7 a), but now with intensities scaled by $|{\bf Q}|^{2}$. At large $|{\bf Q}|$, such as ($-\frac{5}{2},-\frac{5}{2}, 0$) and ($-\frac{7}{2},-\frac{7}{2}, 0$), where phonon scattering dominates all inelastic scattering, the $\chi\prime\prime({\bf Q},\hbar\omega)$ integrals overlap very well, as is expected for phonons. However, at ($-\frac{1}{2},-\frac{1}{2}, 0$), and, to a considerably lesser extent, ($-\frac{3}{2},-\frac{3}{2}, 0$), the data deviates from this $|{\bf Q}|^{2}$ scaling. The scattering at the 2DMZCs correponding to the smallest $|{\bf Q}|$ displays inelastic spectral weight that is much stronger than that expected for phonons alone. From this we conclude that there must be a strong enhancement of the inelastic spectral weight between 15 and 20 meV that is not captured either by one phonon scattering processes or $\hbar\omega ^{-1}$ spin-waves. This analysis does not eliminate the possibility of a purely magnetic effect. That said, a hybrid spin-phonon origin is more plausible, as the enhanced intensity occurs only at coincidences between dispersive spin excitations at the 2DMZCs and non-dispersive phonons. A related effect appears to occur in superconducting LSCO samples\cite{Lipscombe_PRL_2009}. Independent of its origin, the enhancement is clearly a large and significant effect.

\subsection{Comparison to Density Functional Theory and Discussion}

\begin{figure}
\centering 
\includegraphics[scale=0.165]{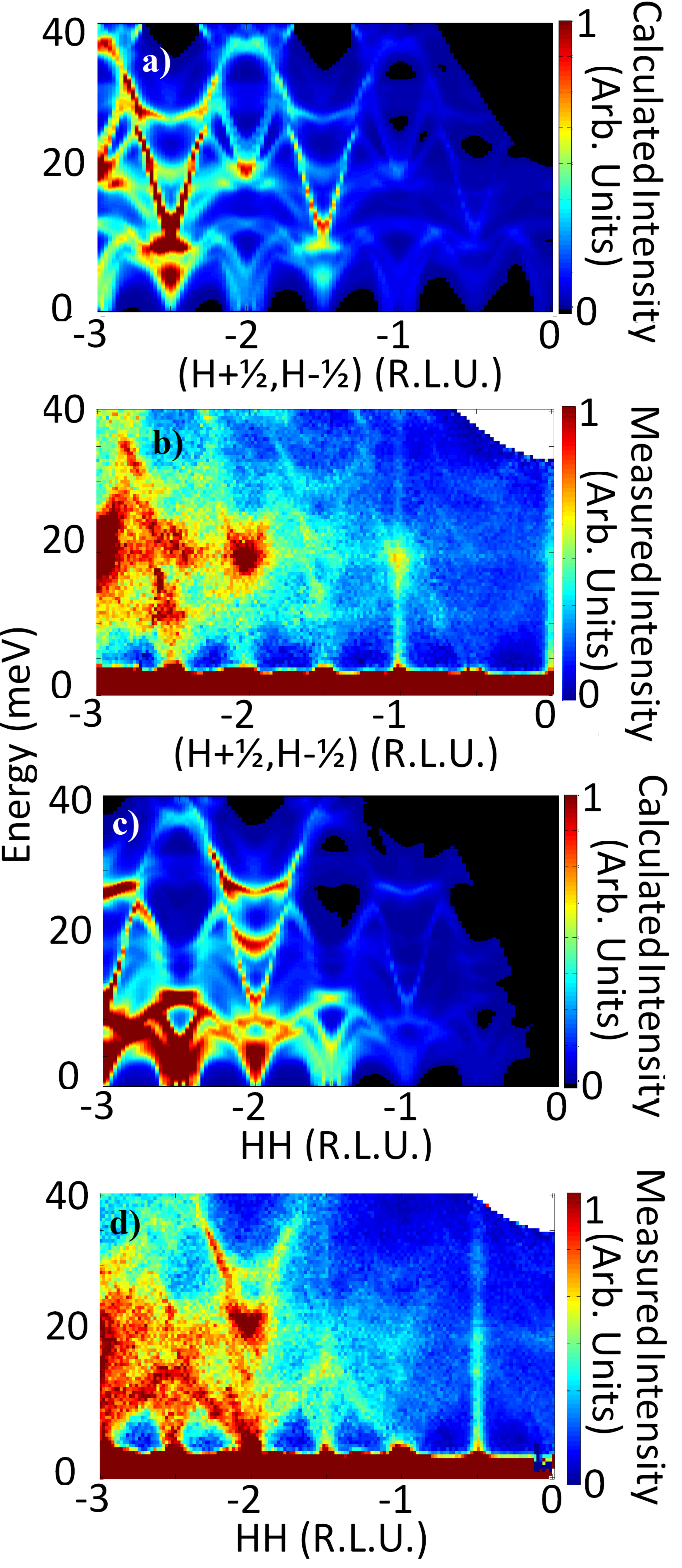} 
\caption{Energy vs. wave-vector maps comparing phonon calculations to the neutron scattering data. All data sets integrate in $L$ from -4 to 4 and $\pm$0.1 about $H\bar{H}$. Panels a) and b) display the dispersion along $(H+\frac{1}{2}, H-\frac{1}{2})$ and integrate about $H\bar{H}$ = 0.5, while panels c) and d) showcase the dispersion along $HH$. Calculations, shown in panels a) and c), display calculated phonon intensities, while panels b) and d) show the corresponding measured data. Panels a) and c) are normalized to the same calculated arbitrary intensity scale while panels b) and d) are normalized to the same measured intensity scale, which is distinct from that used in panels a) and c).}
\label{calc_HHE}
\end{figure}

The analysis pertaining to Figs. 4-7 reveals a consistent picture of robust enhancement of the inelastic spectrum at the low energy spin-phonon crossings. It is therefore important to, at least qualitatively, understand the nature of the relevant phonons involved. To do this, we turn to density functional theory (DFT) appropriate to La$_2$CuO$_4$, whose phonon spectrum should resemble that of LBCO with x = 0.035. Calculations were performed using the density functional perturbation approach as implemented in the mixed-basis pseudopotential framework\cite{Dan_1,Dan_2}. The calculation was performed in the tetragonal structure, using the experimental lattice constants for $(La_{0.7}Sr_{0.3})_{2}CuO_{4}$\cite{Dan_3}. Internal parameters (z-positions of La and $O_{4}$) were optimized to obtain a force-free geometry. The local density approximation was used in the same parametrization employed in Perdew-Wang's work\cite{Dan_4}. The calculated phonon dispersion was obtained by interpolation of dynamical matrices, which were calculated on a 2x2x2 tetragonal mesh. We note that an instability at the M points in the Brillouin zone occurs because the tetragonal structure of pure $La_{2}CuO_{4}$ is not stable at low temperature\cite{Zhao_PRB_2007}.

Typical results for these calculations are shown in Fig. \ref{calc_HHE}, where we compare these calculations to their appropriate neutron scattering counterparts. In Fig. \ref{calc_HHE} a) and c), we show the calculated phonon dispersion and intensities along two parallel wave-vectors: $(H+\frac{1}{2}, H-\frac{1}{2})$ in Fig. \ref{calc_HHE} a) $HH$ in Fig. \ref{calc_HHE} c). These results can be compared with the corresponding neutron scattering data shown in Fig. \ref{calc_HHE} b) and d). All the panels in Fig. \ref{calc_HHE} have the same integration in $<L>$, namely -4$\leq L \leq$4, and both the measured and calculated intensities are shown on full intensity scale. To illustrate the phonon dispersion, we have convolved our calculations with a resolution function that is narrower than the experimental resolution. From this, a rather hard comparison between the measured and calculated phonon spectra can be made.

We note that the DFT calculation does not capture any magnetic scattering. This can be readily seen as dispersive spin excitations emanate from the 2DMZCs in the experiment (Fig. \ref{calc_HHE} b) and d)), but are absent in the calculation (Fig. \ref{calc_HHE} a) and c)). However the DFT calculation does clearly capture optic phonons that are strong near 19 meV, and disperse upwards and away from the 2DMZCs, as is seen in the experiment. The strongest such optic phonon in the field of view for the $HH$ direction shown in Figs. 8 a) and b), appears just above 20 meV at (-3,-3) in both the calculation and the experiment. These comparisons between theory and experiment give us confidence that the DFT calculation is capturing many of the key features in the phonon spectrum for LBCO x = 0.035. We can then use these calculations to understand which optic phonons are participating in the strong enhancement to the intensity that we observe at the 2DMZCs shown in Figs. 5-8. We note that at low energies the comparison between calculation and measurement appears less robust. This is caused by a series of phonon branches whose minima occur at less than 0 meV, a consequence of the proximity to a tetragonal to orthorhombic structural transition in this material.

\begin{figure}
\centering 
\includegraphics[scale=0.17]{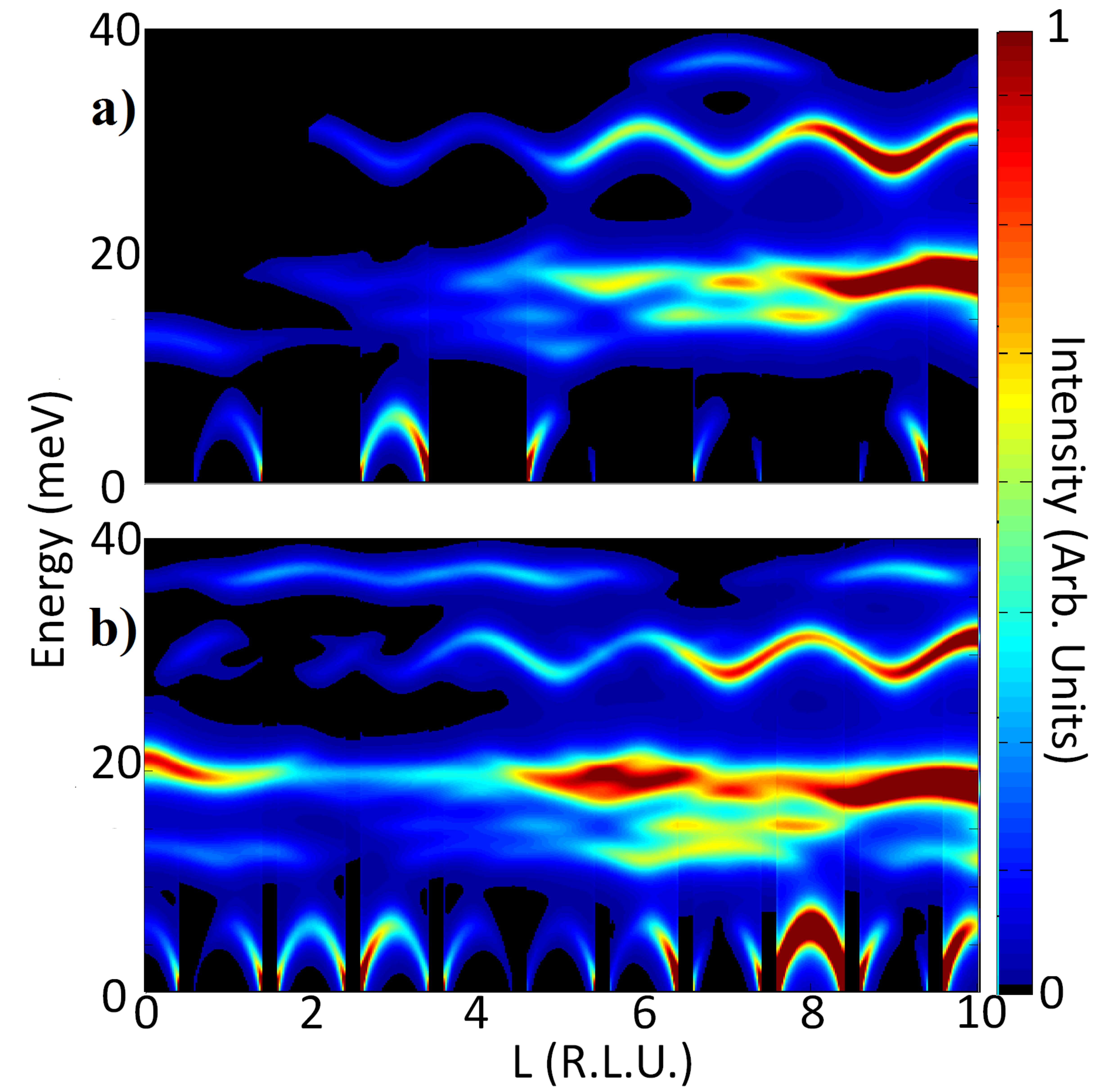} 
\caption{Energy vs $L$ maps of phonon dispersions and intensities from DFT calculations for {\bf Q} = a) ($-\frac{1}{2}, -\frac{1}{2}, L$) and b) ($-\frac{1}{2}, -\frac{3}{2}, L$).}
\label{calc_LE}
\end{figure}

We now consider the energy vs. $L$ dependence of the the calculated phonons at two wavevectors considered in this paper. These are the {\bf Q} = ($-\frac{1}{2},-\frac{1}{2}, L$) and ($-\frac{1}{2},-\frac{3}{2}, L$) 2DMZCs, shown in Fig. \ref{calc_LE} panels a) and b), respectively. At all values of $L$, we observe relatively strong and dispersionless optic phonons near 30, 19 and 15 meV, although the calculated phonon mode with the largest spectral weight and least dispersion is the 19 meV mode. Focusing on the ($-\frac{1}{2},-\frac{1}{2}, L$) 2DZMC, which presents the largest ${\bf Q}^{2}$ normalized enhancements (see Fig. 7), our DFT calculations show that the eigenvector for this $\sim$ 19 meV optic phonon involves atomic displacements that correspond primarily to oxygen displacements within the Cu-O basal plane. Moreover, these displacements do not occur for the oxygen within the La-O layers. Such a phonon eigenvector would be expected to possess a very 2D nature, as the stretching of relatively weak bonds in the third dimension are minimal. A similar case can be made for the nature of the optic phonon near 15 meV and the 2DMZC at ($-\frac{1}{2}, -\frac{1}{2}, L$), although the isolation of the precise eigenvector of the 15 meV optic phonon is less robust than is the case for the 19 meV optic phonon.

These results strongly suggest that the large enhancements in the spectral weight observed at the 2DMZCs and near 15 and 19 meV in LBCO with x = 0.035 result from the confluence of quasi-2D spin fluctuations with quasi-2D optic phonons at the same 2D wave-vectors. As shown in Fig. \ref{small_L}, it is primarily the lowest $|{\bf Q}|$ 2DMZC whose intensity deviates most from the $|{\bf Q}|^{2}$ dependence of the scattered intensity seen at higher $|{\bf Q}|$. Therefore, a possible explanation for the origin of this strong enhancement is hybridization between the quasi-2D spin fluctuations and quasi-2D phonons. Such an interpretation would be natural as the atomic displacements involved in the quasi-2D optic phonons are such that Cu-O-Cu bonds are stretched and distorted. Such distortions affect the nature of the strongest Cu-Cu superexchange pathways in the LBCO system. Independent of the precise origin of this resonant enhancement of the inelastic spectrum in LBCO, it is clear that the energy scale of these effects, which is $\sim$200 K, is large and large enough to play a role in the mechanism underlying high temperature superconductivity which occurs on the same temperature scale. However, any firm connection to superconductivity is lacking at present.

\section{Conclusions}

To conclude, comprehensive time-of-flight neutron scattering measurements have observed rich structure in the energy dependence of the inelastic scattering at 2DMZCs, in particular at those at the lowest-$|{\bf Q}|$, such as ($\frac{1}{2}, \frac{1}{2} L$). This structure presents in the form of strong enhancements of the spectral weight at several crossings of highly dispersive spin excitations with relatively dispersionless phonons in LBCO with x = 0.035. The measured enhancements are large and account for as much as $\sim$25$\%$ of the spectral weight between 5 and 40 meV at the lowest-$|{\bf Q}|$ 2DMZC.

Modeling the phonons in La$_2$CuO$_4$ with density functional theory allowed us to identify the likely eigenvectors associated with the optic phonons involved in the enhancements. These phonon modes appear to be quasi-2D themselves, with appropriate atomic displacements that could affect the strongest Cu-O-Cu superexchange pathways.

This robust structure within the excitation spectrum at the 2DMZCs in LBCO appears on a high energy scale and therefore is of potential relevance to high temperature superconductivity itself. While the current study was carried out on a non-superconducting sample, both the quasi-2D spin fluctuations and the nature of the phonon spectrum for this material family should be slowly varying as a function of doping. We also note that superconducting ground states in LBCO form for x $>$ 0.05, which is only a modest change in x from the present sample. Therefore, the reported energy dependence in the 2DMZC spectral weight, resulting from a confluence of quasi-2D spin fluctuations and quasi-2D optic phonons, is likely a common feature for a broad range of concentrations relevant to superconductivity in the LBCO system. 

\section{Acknowledgements}

We would like to acknowledge useful conversations had with T. Timusk, J. P. Carbotte, I. A. Zaliznyak, J. M. Tranquada, G. E. Granroth, S. A. Kivelson, S. D. Wilson, N. B. Christensen, J. Gaudet, B. Jackel and J. L. Niedziela. Research at ORNL's Spallation Neutron Source was sponsored by the Scientific User Facilities Division, Office of Basic Energy Sciences, U.S. Department of Energy. This work was supported by NSERC of Canada.

\bibliographystyle{unsrt}
\bibliography{cuprate}

\end{document}


\title{Supplemental Material: Quasi-Two Dimensional Spin and Phonon Excitations in $La_{1.965}Ba_{0.035}CuO_{4}$}

\maketitle

\section{Background Subtraction}

We first discuss the measurement of the empty sample can background for the inelastic neutron scattering data presented in the main manuscript, its origin and magnitude. We explicitly show that this background constitutes a small fraction, $\sim$ 2-3 $\%$ or less, of the total coherent neutron scattering at small $|{\bf Q}|$, and thus can readily be corrected for by subtracting an empty sample can data set, from the ``signal" data sets with the sample in place. 

The determination of the background is generally important for any scattering experiment\cite{Tranquada_Text}. It can display structure as a function of $|{\bf Q}|$ and energy, or both, which can mimic scattering from the sample. It can also contribute to multiple scattering, if it is sufficiently strong. Indeed, this latter effect has been the subject of recent neutron scattering measurements\cite{Pintschovius_Arxiv_2014}. Moreover, as discussed in the main paper, the isolation of $\chi\prime\prime$({\bf Q},$\hbar\omega)$ from the measured S({\bf Q},$\hbar\omega$) requires that the background be well determined such that the Bose thermal population is applied only to the signal and not to the background. Consequently, understanding and accounting for background is important to the quantitative analysis that we performed here and is common to many inelastic neutron scattering studies.

\begin{figure}[b]
\renewcommand{\thefigure}{S\arabic{figure}}
\centering 
\includegraphics[scale=0.16]{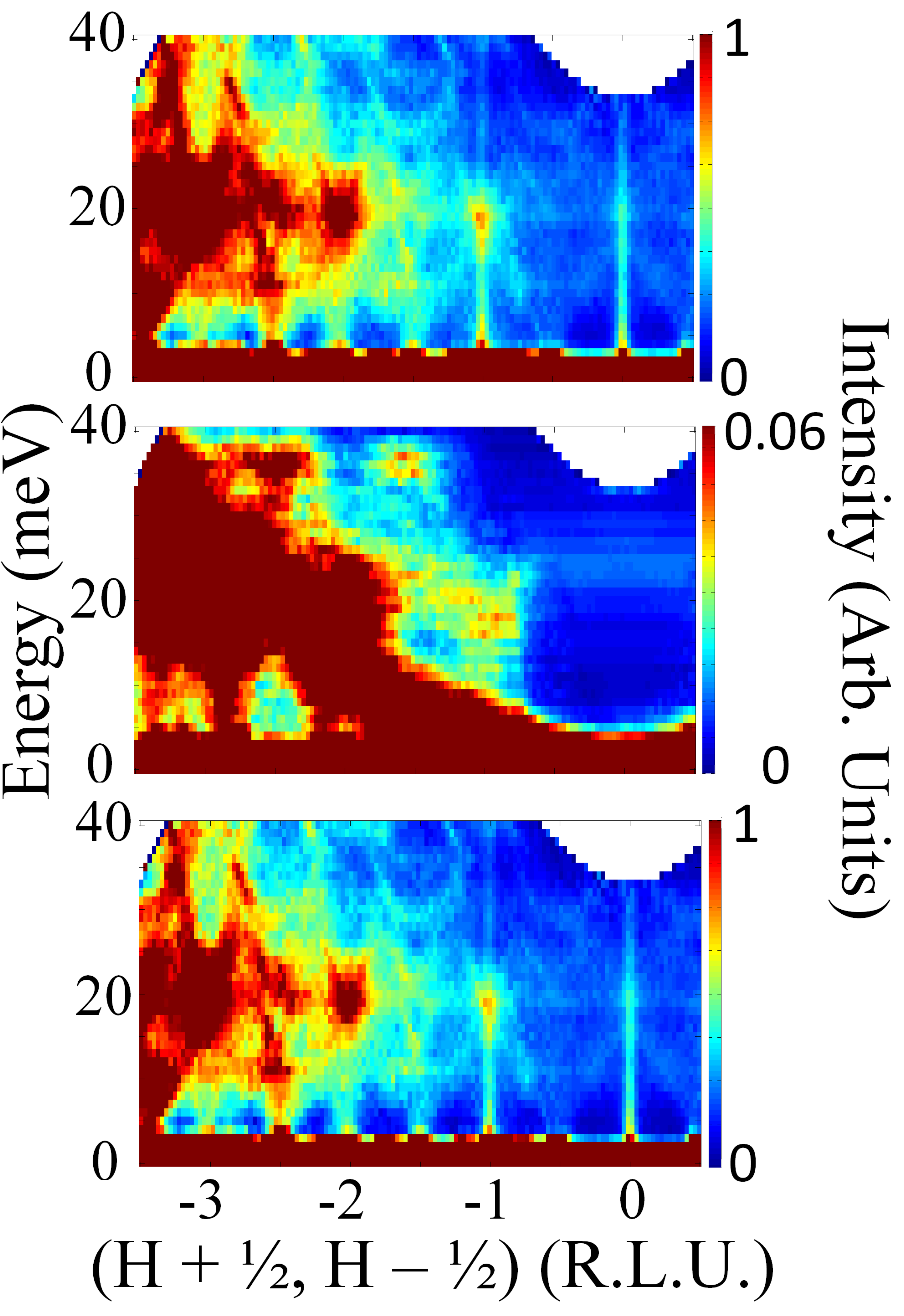} 
\caption{Energy vs. wave-vector maps produced nearly identically to Fig. 1 of the main manuscript. The only distinction here is that instead of integrating from 0.4 to 0.6 in $\bar{H}H$, as was done in Fig. 1 of the main manuscript, here we integrate from -0.1 to 0.1. a) A normalized signal data set with no empty can subtraction is shown. b) A normalized empty sample can data set on an intensity scale 6$\%$ of that of either a) or c) is shown. c) A normalized signal minus background data set is shown on the same intensity scale as in a). This is the same data as presented in Fig. 1 of the main manuscript. All data sets shown were collected at 7 K.}
\end{figure}

\begin{figure}
\renewcommand{\thefigure}{S\arabic{figure}}
\centering 
\includegraphics[scale=0.18]{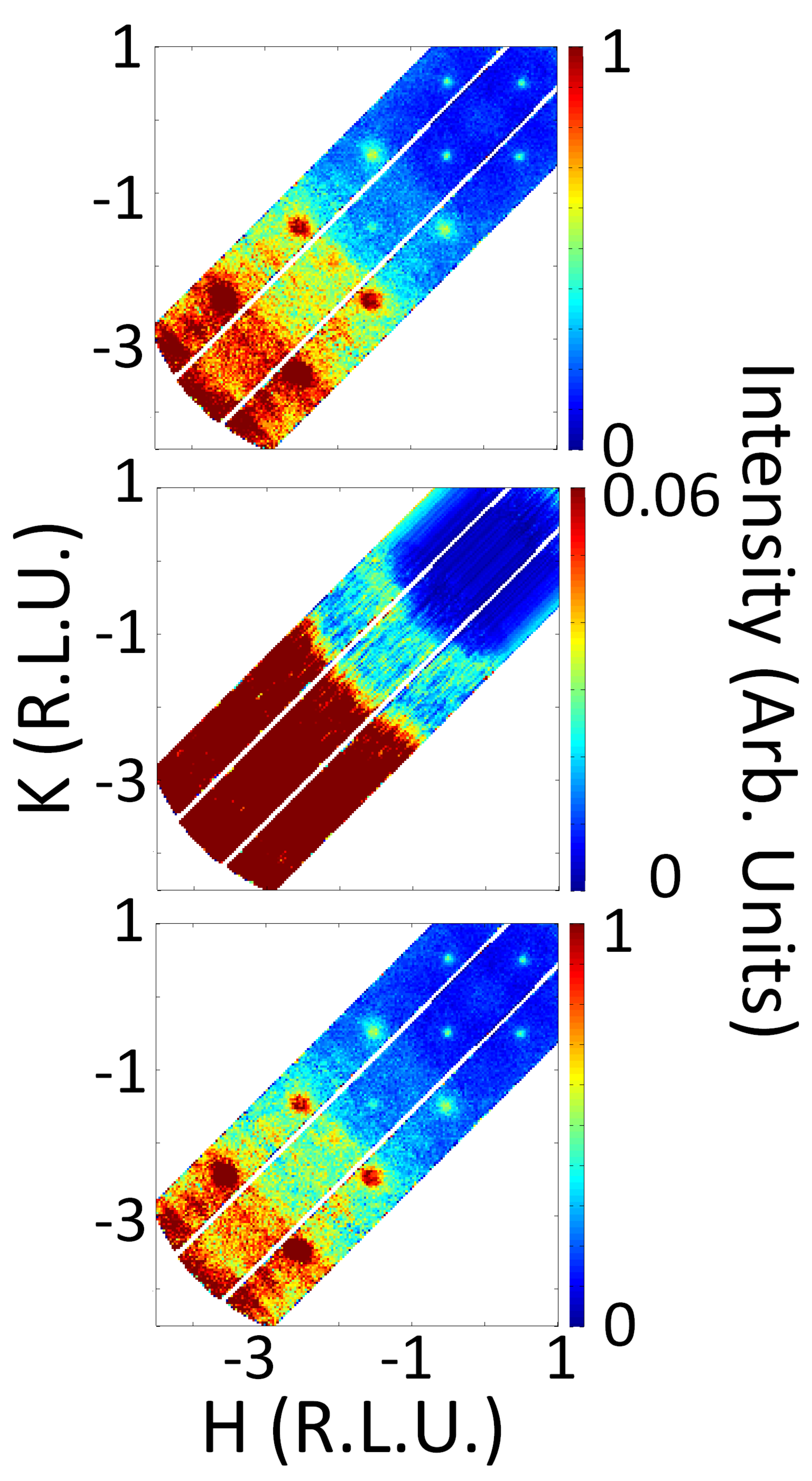} 
\caption{$HK$ maps produced identically to Fig. 2 of the main manuscript. a) A normalized signal data set with no empty can subtraction is shown. b) A normalized empty sample can data set on an intensity scale 6$\%$ of that of either a) or c) is shown. c) A normalized signal minus background data set is shown on the same intensity scale as in a). This is the same data as presented in Fig. 2 of the main manuscript. All data sets shown were collected at 7 K.}
\end{figure}

\begin{figure}[tbp]
\renewcommand{\thefigure}{S\arabic{figure}}
\centering 
\includegraphics[scale=0.17]{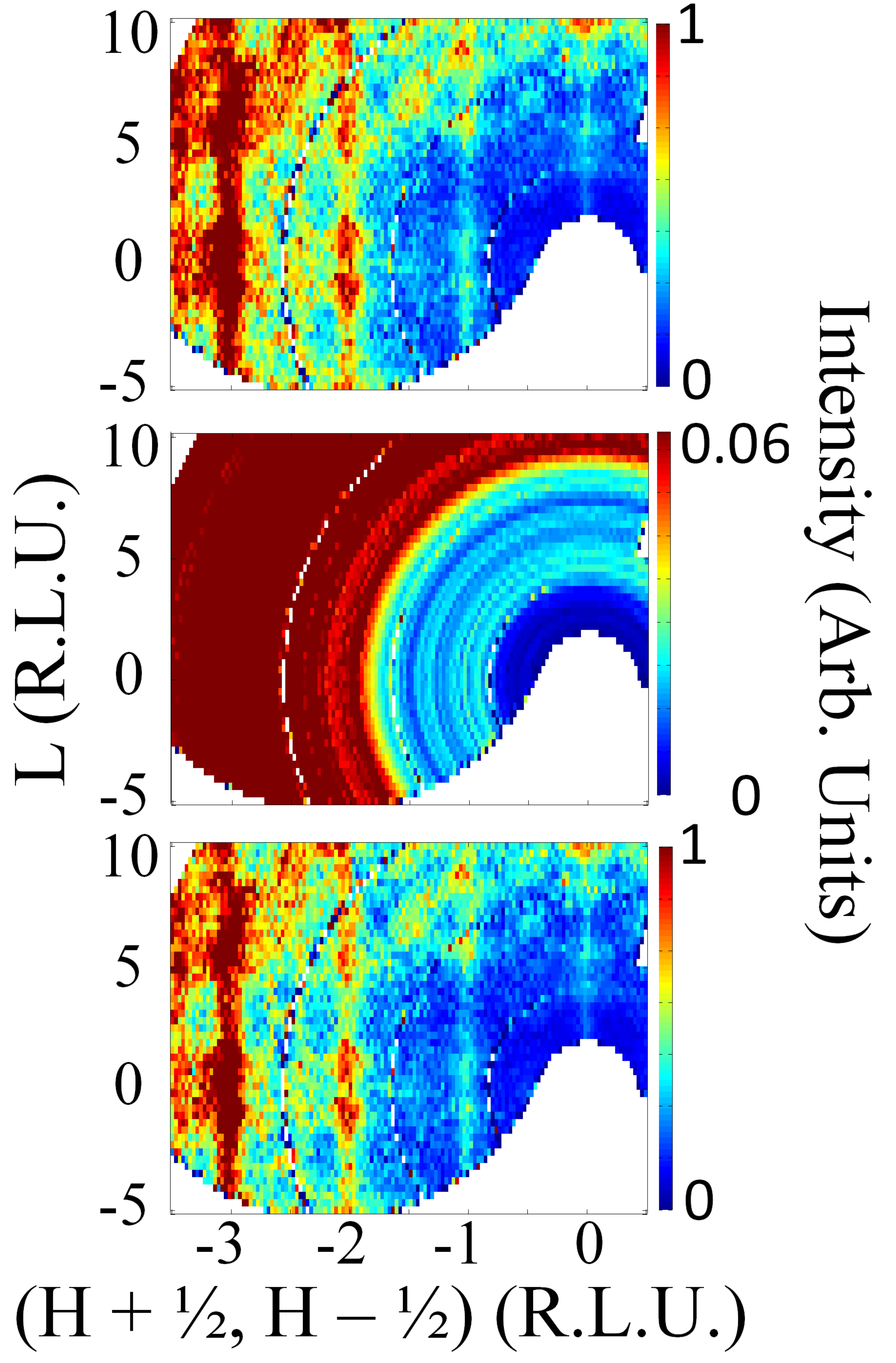} 
\caption{$HHL$ maps produced identically to Fig. 3 of the main manuscript. a) A normalized signal data set with no empty can subtraction is shown. b) A normalized empty sample can data set on an intensity scale 6$\%$ of that of either a) or c) is shown. c) A normalized signal minus background data set is shown on the same intensity scale as in a). This is the same data as presented in Fig. 3 of the main manuscript. All data sets shown were collected at 7 K.}
\end{figure}

In most neutron scattering experiments, and certainly in the present experiment, the primary source of background is scattering from the adenda, which are comprised from the mount for the single crystal, any radiation shields that are in place and other parts of the cryostat that may be in the neutron beam. These adenda are made from machined aluminum, which is used because Al has a low coherent and incoherent neutron cross sections, has little neutron absorbtion and is easily machined. The neutron scattering from the adenda is therefore expected to be weak. 

Despite the weak contributions of the adenda, care is taken to minimize the amount of material from the adenda in the neutron beam and also to shield as much of this remaining material as possible by using strong neutron absorbers. In the present experiment, we employed a cadmium mask of the sample mount. ``Empty can" measurements are then collected, which are scans performed without the sample in place but with the adenda still present. Such scans then serve as a measurement of the background and it is these empty can data sets that are subtracted from the those performed with the aligned sample in place. All the data presented in the main manuscript is this background-corrected data. Here we demonstrate the relative strength of the measured background to the measured signal.

Figures S1, S2, and S3 show Energy vs. $HH$, $K$ vs. $H$, and $L$ vs. $HH$ maps of the inelastic neutron scattering from our $La_{1.965}Ba_{0.035}CuO_{4}$ experiment, with the scattered neutron intensity shown on full scale. We note that Figs. S1 c), S2 c) and S3 c) are the same empty can background-corrected data sets that were presented and discussed in the main manuscript. Figs. S1 a), S2 a) and S3 a) show the original raw data, without subtraction of the empty can data, displayed on the same intensity scale as S1 c), S2 c), and S3 c). Figs. S1 b), S2 b), and S3 b) show the results from the ``empty can" measurements alone and are presented on a full intensity scale that is only 6 $\%$ of that of the ``signal" scans. This clearly shows that the measured background is a small correction to the scattered intensity from the sample, especially at small $|{\bf Q}|$. We therefore conclude that our background subtraction of these data sets is robust, and multiple scattering involving the adenda is not a significant issue for this experiment, especially at low $|{\bf Q}|$.

\section{Energy Dependent Background Analysis}

Once the ``empty can" background has been removed from a data set, one may still wish to isolate signal of a specific origin, such as magnetic scattering, which has a specific wave-vector dependence, from another signal, such as phonon scattering, with a different wavevector dependence. Below we present an alternate approach to estimating the scattered intensity that occurs at the 2DMZCs to that presented in the main manuscript. With this approach, we use the intensity measured within the basal plane away from the 2DMZCs as an energy-dependent background. This energy dependent background method accounts for phonon contributions with little dispersion and $|{\bf Q}|$ dependence to their intensity. As in the main manuscript, in using this method we employ an additional set of integrations, which, instead of displaying slices of the data, as those shown in Figs. 1-3 of the main manuscript, yield effective constant energy cuts through the data.

\begin{figure}
\renewcommand{\thefigure}{S\arabic{figure}}
\centering 
\includegraphics[scale=0.15]{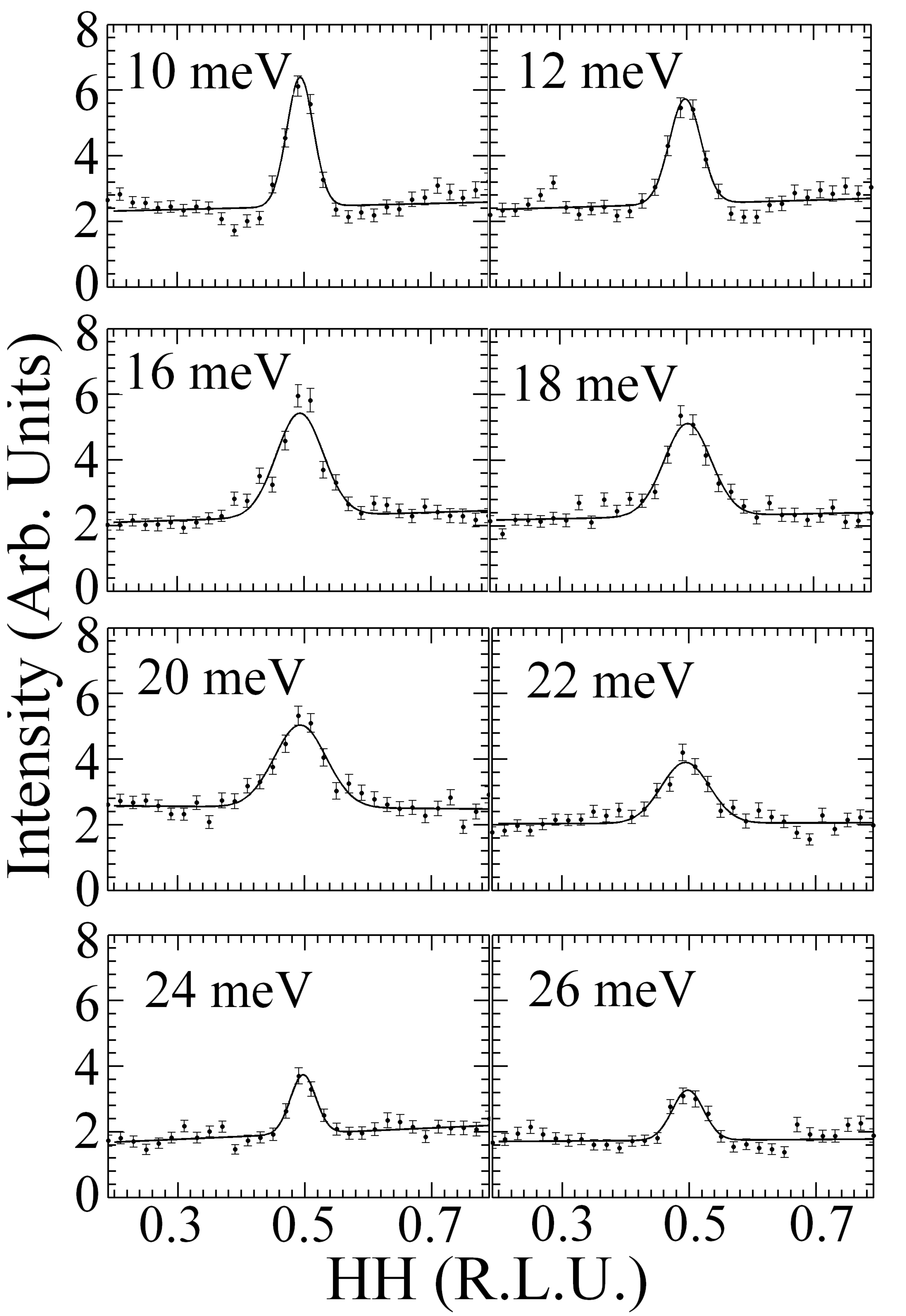} 
\caption{Evolution of the integrated inelastic neutron scattering intensity near ($\frac{1}{2},\frac{1}{2}$) as a function of energy as observed in a series of constant energy cuts integrating -4 to 4 in $<L>$, -0.1 to 0.1 in $<H\bar{H}>$ and $\pm$ 1 meV in energy. Solid lines are fits of the data, where the peaks are modeled as Gaussians and the background is modeled phenomenologically as a sloping linear background. Error bars represent one standard deviation.}
\end{figure}

\begin{figure}
\renewcommand{\thefigure}{S\arabic{figure}}
\centering 
\includegraphics[scale=0.13]{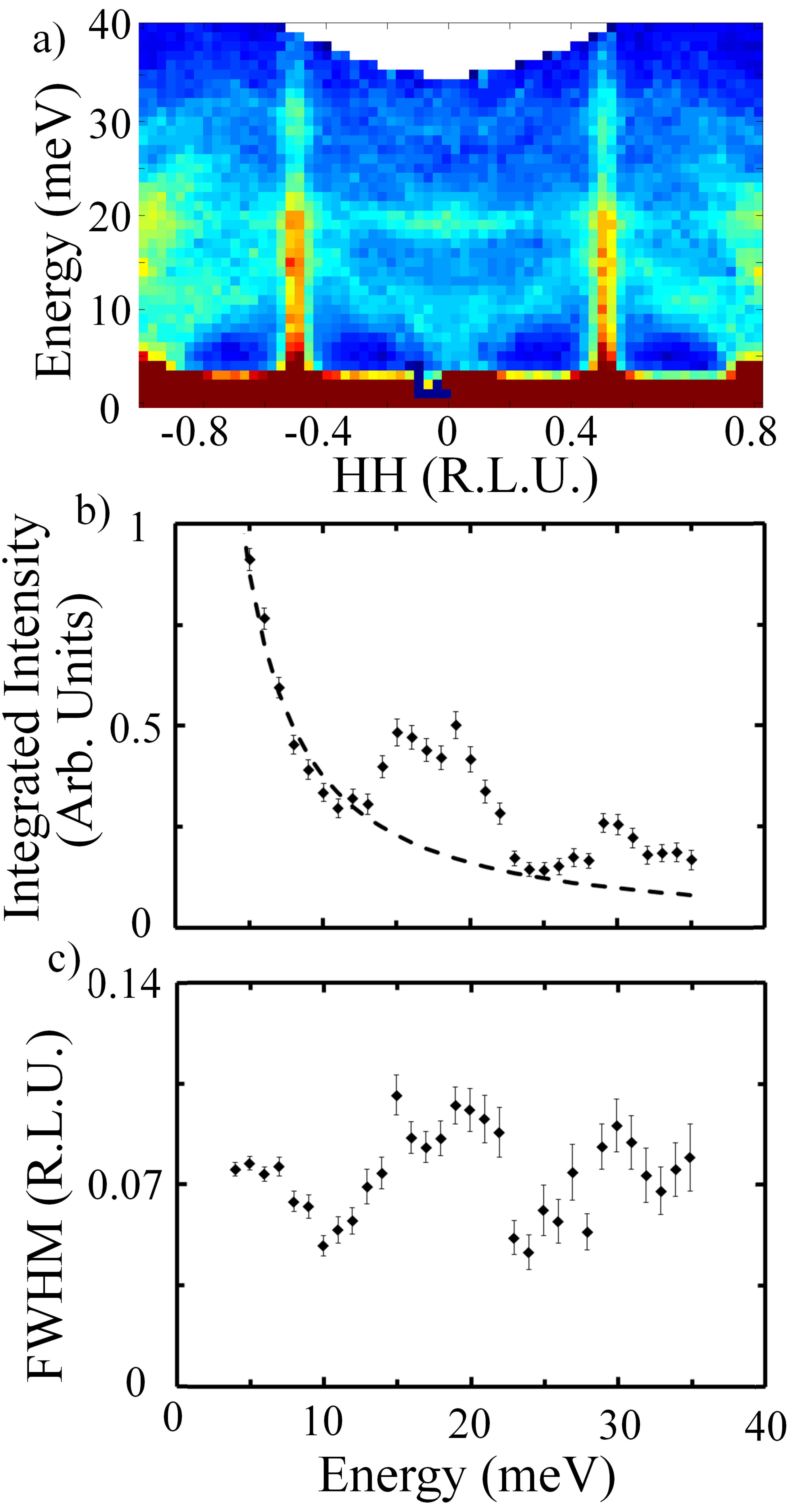} 
\caption{a) An expanded view of the ($\frac{1}{2},\frac{1}{2}$) and ($-\frac{1}{2},-\frac{1}{2}$) 2DMZCs, integrating -4 to 4 in $L$, and -0.1 to 0.1 in $H\bar{H}$. This contour map again displays the rich structure that develops at the crossings of the dispersive spin excitations with the phonons in this system. b) Resultant integrated intensity at the 2DMZC ($\frac{1}{2}$,$\frac{1}{2}$) positions from phenomenological modelling. The integrated intensity is the area of the Gaussians employed in the fits, as shown in Fig. S4. The dashed curve is a phenomenological curve of best fit, which we find to go as a E$^{-1.2}$, likely due to the large $L$ integration we use here. This curve is suggestive of the energy dependence resultant from spin excitations alone and serves to demonstrate the dramatic nature of the enhancement of the scattering at the spin-phonon crossings between 10 and 40 meV. c) Corresponding full width half maximum (FWHM) obtained from the Gaussian fits of the data from Fig. S4. These give the FWHM as measured along the $HH$ direction. In this case, the integration range reflects that we have employed an energy, $L$ and $H\bar{H}$ integration, as described in the the text, in addition a fit along $HH$. Error bars represent one standard deviation.}
\end{figure}

We now consider such constant energy cuts along $HH$ through and in the vicinity of ($\frac{1}{2},\frac{1}{2}$), which is achieved by integrating in energy, $H\bar{H}$ and $L$. We chose this position in reciprocal space as it is the 2DMZC with the smallest $|{\bf Q}|$, implying that it has minimal contributions to the scattered intensity from phonons and maximal contributions from purely magnetic scattering. We display a series of such representative cuts from 10 to 26 meV in Fig. S4. The solid lines in this figure are fits using a Gaussian centered at ($\frac{1}{2}$,$\frac{1}{2}$) to represent the scattering at the 2DMZC plus a linear background, which is a phenomenological fit for contributions from relatively dispersionless phonons. As can be seen in Fig. S4, the background is largely linear in $HH$, although it can display somewhat more complex wavevector structure. Nonetheless, we can phenomenologically model these trends and thereby isolate the nominal scattering centered on the ($\frac{1}{2},\frac{1}{2}$) 2DMZC position. 

Figure S5 a) shows an energy vs. $< HH >$ wave-vector map featuring the ($\frac{1}{2},\frac{1}{2}$) and (-$\frac{1}{2},-\frac{1}{2}$) 2DMZC positions. This plot is similar to that shown in Fig. 1 of the main manuscript, however there we presented the inelastic scattering spectrum through 4 2DMZCs, as a function of $< H+1/2, H-1/2 >$ with systematically decreasing $|{\bf Q}|$ from left to right. We note that the reciprocal space direction shown in Fig. 1 of the main manuscript is parallel to $<HH>$. Here, Fig. S5 a) again highlights the structure that develops at several of the spin-phonon dispersion crossings in this system. 

Figure S5 b) displays the resulting energy dependence of the integrated scattered intensity centered on ($\frac{1}{2},\frac{1}{2}$), which we take to be the area of the Gaussian fits in Fig. S4. It shows the ($\frac{1}{2},\frac{1}{2}$) scattering to fall dramatically between 5-10 meV before rising by factors of 3-4 between 15-20 meV. For context, the characteristic energy dependence of the integrated intensity for spin waves is expected to diminish as $\frac{1}{\hbar\omega}$.\cite{Tranquada_Text} The dashed line in Fig. S5 b) is the result of modeling the scattering away from any spin-phonon crossings by a phenomenological form of E$^{-1.2}$. This model provides a good description of the data at low energies and is suggestive of the energy dependence from spin excitations that exist away from the crossings with the various phonons. Fig. S5 b) demonstrates, therefore, the surprisingly large scale of this enhancement. This enhancement occurs at all three spin-phonon crossings within this field of view: at $\sim$ 15 meV, $\sim$ 19 meV and $\sim$ 30 meV, which is where the low energy and relatively dispersionless optic phonon bands cross the 2DMZC. A simple estimate of the scattered intensity due to this enhancement suggests that at ($\frac{1}{2},\frac{1}{2}$) approximately 25$\%$ of the total scattered intensity below 40 meV arises from these spin-phonon crossings.

Figure S5 c) shows the corresponding full width half maximum (FWHM) obtained from the Gaussian fits of the scattering at the 2DMZC. Note that the FWHM first deceases by almost of factor of 2 between 5 and $\sim$10 meV, which corresponds to the hour-glass dispersion of the 2D incommensurate spin excitations. This is consistent with earlier work on the LSCO system, which shows that the ``waist" of the hour-glass occurs around $\sim$11 meV for a doping of x = 0.035\cite{Matsuda_PRL_2008}. The FWHM also shows pronounced peaks at each of the three spin-phonon crossings below $\sim$40 meV. 

We note that there is a high degree of consistency between the results of this analysis and that performed using an energy-independent background subtraction, as presented in Figs. 5 and 6 of the main manuscript. 

\bibliographystyle{unsrt}
\bibliography{cuprate}